\DocumentMetadata{}
\documentclass[sigconf,nonacm]{acmart}

\AtBeginDocument{%
  \providecommand\BibTeX{{%
    \normalfont B\kern-0.5em{\scshape i\kern-0.25em b}\kern-0.8em\TeX}}}

\copyrightyear{2024}
\usepackage{subcaption}
\usepackage{stfloats} 
\begin{document}

\title[Leveraging Large Language Models for Realizing Truly Intelligent User Interfaces]{Leveraging Large Language Models for Realizing Truly Intelligent User Interfaces}

\author{Allard Oelen}
\email{allard.oelen@tib.eu}
\orcid{0000-0001-9924-9153}
\affiliation{
  \institution{TIB – Leibniz Information Centre for Science and Technology}
  \city{Hannover}
  \country{Germany}
}

\author{Sören Auer}
\email{soeren.auer@tib.eu}
\orcid{0000-0002-0698-2864}
\affiliation{
  \institution{TIB – Leibniz Information Centre for Science and Technology}
  \city{Hannover}
  \country{Germany}
}

\renewcommand{\shortauthors}{Oelen and Auer}

\begin{abstract}
The number of published scholarly articles is growing at a significant rate, making scholarly knowledge organization increasingly important. 
Various approaches have been proposed to organize scholarly information, including describing scholarly knowledge semantically leveraging knowledge graphs. 
Transforming unstructured knowledge, presented within articles, to structured and semantically represented knowledge generally requires human intelligence and labor since natural language processing methods alone typically do not render sufficient precision and recall for many applications. 
With the recent developments of Large Language Models (LLMs), it becomes increasingly possible to provide truly intelligent user interfaces guiding humans in the transformation process. 
We present an approach to integrate non-intrusive LLMs guidance into existing user interfaces.
More specifically, we integrate LLM-supported user interface components into an existing scholarly knowledge infrastructure. 
Additionally, we provide our experiences with LLM integration, detailing best practices and obstacles. 
Finally, we evaluate the approach using a small-scale user evaluation with domain experts. 
\end{abstract}


\maketitle

\section{Introduction}
Scholarly knowledge organization becomes more important as the rate at which scholarly articles are being published keeps increasing~\cite{Jinha2010}. A few approaches focus on organizing scholarly knowledge in a semantic manner using knowledge graphs. Knowledge graphs are defined as networks of entities, properties, and relationship between entities~\cite{kroetsch2016special}. A common method to describe statements in knowledge graphs is formulating them using a \textit{subject}, \textit{predicate}, and \textit{object}~\cite{powers2003practical} in the Resource Description Framework (RDF)~\cite{manola2004rdf}. RDF adopts terminology from linguistics. The task of transforming scholarly knowledge from an unstructured textual format to a structured format is cumbersome, as it requires both data modeling skills and domain knowledge. The Open Research Knowledge Graph (ORKG)~\cite{auer2020improving} aims to create such a knowledge graph containing knowledge presented in scholarly articles. The ORKG leverages crowdsourcing to perform the knowledge transformation process, by including article authors in this process~\cite{10.1145/3397481.3450685}. The ORKG is a domain-agnostic knowledge graph, targeting a variety of scholarly domains. Due to the domain-agnostic nature, knowledge is not described by a predefined ontology, but is rather self-evolving based on domain requirements. Data models are crowdsourced by domain experts, which are materialized in ORKG Templates. Templates have a similar function as SHACL shapes~\cite{corman2018semantics} in traditional RDF knowledge graphs. Since users are able to create new ontology terms within the graph, system guidance is of importance to ensure a correct and complete structured article description. In light of the recent advancements of Natural Language Processing (NLP) and more specifically of Large Language Models (LLMs)~\cite{radford2018improving}, it becomes increasingly possible to use Artificial Intelligence (AI) to assist users in the transformation process. Particularly, ChatGPT gained significant attention across a plethora of domains, including health care~\cite{biswas2023role}, scientific publishing~\cite{hill2023chat}, and programming~\cite{surameery2023use}.

\begin{figure*}
\centering
\begin{subfigure}[t]{\columnwidth}
  \centering
  \includegraphics[width=\linewidth]{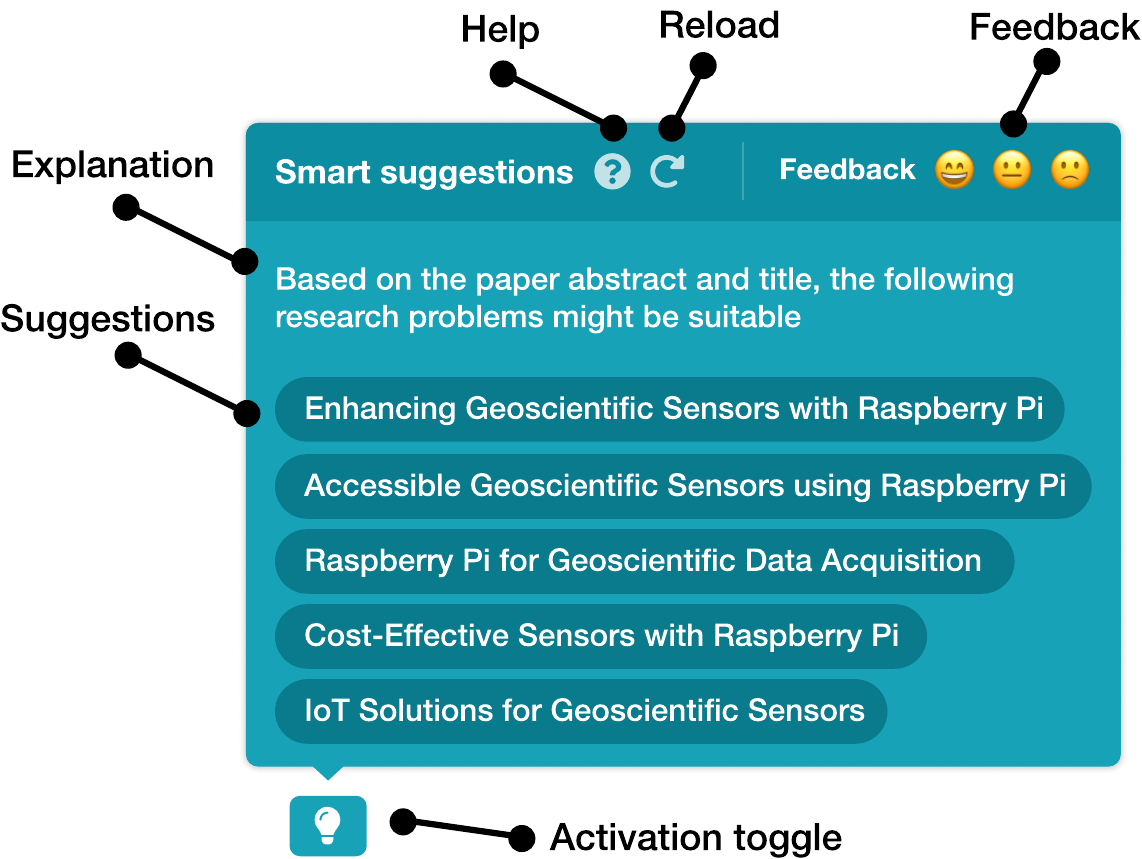}
  \caption{The Smart Suggestion tooltip is a reusable UI component. The content changes dynamically based on the specific use case. }
  \Description{Screenshot of the UI. An activation toggle is displayed as a button with light bulb icon. An open tooltip is displayed which is the Smart Suggestion. The tooltip uses a distinctive blue color to communicate to users that the component is supported by AI (in addition to the Smart Suggestions text). Furthermore, a use case is displayed there the LLM recommend 5 different values for the 'research problem' predicate. }
  \label{fig:implementation-1}
\end{subfigure}\hfill
\hspace{.5em}\begin{subfigure}[t]{\columnwidth}
  \centering
  \includegraphics[width=.8\linewidth]{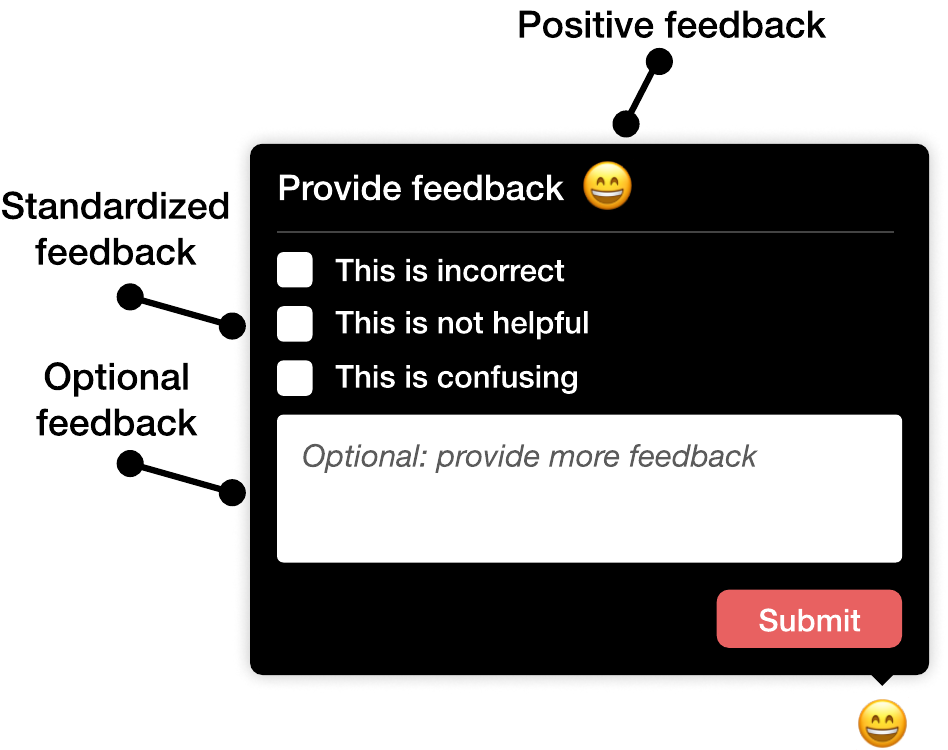}
  \caption{For each LLM use case, user feedback can be provided, which in turn is used to improve the system.}
  \Description{A screenshot of the feedback tooltip that can be opened from within the Smart Suggestions. A list of three detail options are provided, where users can indicate is the suggestion is helpful, correct, or confusing. Furthermore, a text area is displayed where they can optionally provide additional feedback. }
  \label{fig:implementation-2}
\end{subfigure}
\caption{The two main components of the UI implementation.}
\label{fig:implementation}
\end{figure*}

In this work, we propose an approach to integrating LLMs into existing User Interfaces (UIs). Implementing LLMs into existing UIs has several implications for example, the integration should be non-intrusive to ensure the existing workflows are not disrupted. Compared to creating completely new UIs using LLMs, the integration into existing systems entails additional challenges and constraints. We present an approach that provides best practices when implementing LLMs into existing UIs. To demonstrate the usefulness of this approach, we use the guidelines to implement LLM support into the ORKG UI. This approach resembles a neuro-symbolic hybrid approach, where the intelligence of neural models (LLM) is combined with semantic representations (knowledge graph) and human intelligence for validation and curation. The LLM is implemented in a UI component called Smart Suggestions. The Smart Suggestions provide guidance to users for a selected set of tasks that users frequently perform when interacting with the ORKG UI. 

In this work, we discuss our experiences with integrating LLMs into UIs by presenting a use case for the scholarly knowledge organization domain. More specifically, we make the following contributions: 
1) We present an approach comprising 22 guidelines for LLM integration into existing UIs.
2) We implement the approach in the ORKG UI by means of Smart Suggestions.
3) We evaluate the approach and implementation with a small-scale user study.

\begin{table*}[]
\centering
\caption{Approach for integrating LLM assistance into existing UIs. Guidelines are formulated as requirements to facilitate implementation. The asterisks * denotes functional requirements, other remaining items resemble non-functional requirements.}
\label{tab:approach}
\resizebox{1\textwidth}{!}{%
\begin{tabular}{@{}p{3.45cm}|p{6.75cm}|p{7cm}@{}}
\toprule
\textbf{Task} & \textbf{Description} & \textbf{Implementation directions} \\ \midrule
\multicolumn{3}{@{}p{\textwidth}@{}}{\textbf{1. Transparency}} \\ \midrule
1.1. Distinguishable & The system shall be clearly distinguishable in the UI. & Using distinctive color scheme and recognizable icons. \\
1.2. Suggestions & The system shall be displayed explicitly as suggestive. & Informing users that the suggestions can be wrong or misleading. \\
1.3. Transparency & The system shall make it clear how suggestions are generated. & Mention the model (e.g., ChatGPT), model input, and prompt. \\
1.4. Multiple variants* & The system shall provide multiple values when appropriate to stress uncertainty. & Provide a list of different options from which users have to select the desired option. \\
1.5. Language & The system shall use appropriate language to express uncertainty. & Use words such as might, could, possibly, seems to be, etc. \\ \midrule

\multicolumn{3}{@{}p{\textwidth}@{}}{\textbf{2. Control}} \\ \midrule
2.1. Non-intrusive* & The system shall have the option to hide it. & Use collapsible UI components. \\
2.2. On demand* & The system shall be displayed on demand. & Do not open the suggestions by default.  \\
2.3. Deactivation* & The system shall provide an option to be deactivated. & Provide a setting on user level to hide the suggestions in the entire UI.
\\ \midrule

\multicolumn{3}{@{}p{\textwidth}@{}}{\textbf{3. Usability}} \\ \midrule
3.1. UI integration & The system shall seamlessly blend into the UI. & Instead of a separate UI, integrate the LLMs into the existing UIs, ensuring the users' attention is focused towards the task. \\
3.2. Consistent availability* & The system shall be available when expected by users. & Smart Suggestions should be available both when adding and editing data. \\
3.3. Optional usage & The system shall not be required to fulfill the task. & Users can still perform the task manually. \\
3.4. Modifiable* & The system shall provide the option to modify suggestions. & After selecting a recommended value, allow the possibility to edit the value. \\
3.5. Regenerating* & The system shall provide an option to regenerate the response. & Using a reload button to get additional LLM responses. \\ \midrule

\multicolumn{3}{@{}p{\textwidth}@{}}{\textbf{4. Error Management}} \\ \midrule
4.1. Graceful degradation & The system shall not break the UI when it is failing. & In case the LLM is not available or not returning the response as expected, ensure the UI remains operable and do not present them as critical errors. \\
4.2. Error recovery* & The system shall provide a possibility to recover from errors. & Add a reload button when errors appear and explain how to present errors. \\
4.3. Error prevention & The system shall minimize the user input to mitigate potential errors. & Prevent errors by built-in prompts with placeholders that contain user input. \\ \midrule

\multicolumn{3}{@{}p{\textwidth}@{}}{\textbf{5. Feedback and Statistics}} \\ \midrule
5.1. High-level feedback* & The system shall facilitate the process of providing feedback with minimal effort. & A three-level scale: positive, neutral, negative to determine whether tasks are performing well, need to be improved, or need to be removed. \\
5.2. Detailed feedback* & The system shall facilitate the process of providing more detailed feedback. & Standardized answers to indicate usefulness and correctness. Optionally provide additional input. \\
5.3. Usage statistics & The system shall be recording usage statistics without explicit efforts from users. & Record clicks when LLM suggestions are being used. \\ \midrule

\multicolumn{3}{@{}p{\textwidth}@{}}{\textbf{6. System Performance}} \\ \midrule
6.1. Response time & The system shall respond within seconds. &  Ensure prompts and answers are short to ensure users can access the LLM tool to get quick access.\\
6.2. Minimize requests & The system shall debounce function calls to minimize requests for environmental and monetary reasons. & Activate LLM support on demand when a button is clicked. \\
6.3. Prevent misuse &  The system shall use a backend service to generate the prompts being sent to the LLM. & Prompts are stored in the service and the LLM interface is not exposed to the client, but made available through middleware. \\
\midrule

\end{tabular}%
}
\end{table*}

\section{Approach and Implementation}
To introduce our approach, we first present guidelines for non-intrusive LLM integration into existing user interfaces. Afterwards, we explain how these guidelines are implemented in a scholarly knowledge infrastructure, by discussing use cases and implementation details. 

\subsection{Guidelines}
\label{section:approach}
To support our key objectives of LLM integration into user interfaces, we formulated a set of 22 guidelines that form the six pillars of our approach. These guidelines are grounded on the ten usability heuristics by \citeauthor{10.1145/191666.191729}~\cite{10.1145/191666.191729} and we further developed them to provide more actionable guidelines specifically targeting the integration of LLMs into UIs. They provide practical guidance while keeping usability in mind. The guidelines revolve around transparency towards users, making it clear where suggestions are coming from, and clearly stating that they might be incorrect. Furthermore, positioning the LLM integration as optional and non-intrusive ensures that users are able to only use LLM support when deemed helpful, and can thus be ignored when this is not the case. This specifically facilitates the adoption of LLMs in existing UIs, even when suggestions by the LLM are not always accurate or helpful. 

To increase the practical usefulness of the guidelines, we formulated them as implementable system requirements. To structure the requirements, we adopted the Easy Approach to Requirements Syntax (EARS) method~\cite{5328509} which provides constraints for how the requirements should be formulated. We discuss the pillars and their system requirements in more detail in \autoref{tab:approach}. When using the term ``the system'', we refer to the implementation of LLMs into a UI. For each guideline, in addition to the description, directions for system implementation are provided.

\subsection{Requirements Implementation}
The proposed approach is implemented into an established scholarly organization platform. Given that one of the main objectives of our approach is to implement LLMs into existing workflows, we did not alter existing user processes. We will now discuss how the LLM support is integrated in six different use cases, and finally provide additional technical details. We used the requirements formulated in \autoref{section:approach} while implementing LLM support into the ORKG UI. A screenshot of the implementation is displayed in \autoref{fig:implementation}. In the UI, a light bulb icon is displayed next to components that support Smart Suggestions. Also, the color blue scheme is recognizable and specifically used for Smart Suggestions.

\subsection{Use Cases}
\label{section:use-cases}
We implemented Smart Suggestions for six different use cases in the ORKG UI. As the use cases are highly domain specific, we will now discuss without going into specifics. The use cases can be categorized into two groups, ``Closed Recommendations'' and ``Open Feedback''. The first category provides the users with a fixed set of different actionable options. The example displayed in \autoref{fig:implementation-1} falls into this category. This specific example provides relevant research problems (i.e., RDF objects) for a specific paper, based on the title and abstract. The second category provides textual feedback to a user, which helps users with the task at hand. In this case, the suggestions displayed in \autoref{fig:implementation-1} are replaced with a paragraph of textual feedback. Most of the use cases that display textual feedback, provide modeling advice to users. For example, they will explain to users whether the data type is chosen correctly (e.g., for text, also called literals in RDF), or by recommending how to formulate labels to enhance data reusability. The individual use cases are discussed in detail in \autoref{appendix:use-cases}. 

\subsection{Technical Details}
\label{section:implementation-technical-details}
One of the key implementation strategies is to make the Smart Suggestions modular and therefore easy to extend. The ORKG infrastructure separates the code base into a frontend, responsible for the UI, and a backend part. The Smart Suggestions feature is implemented in the frontend using the React\footnote{\url{https://react.dev}} framework and is written in JavaScript. Since React uses a component-based approach, we created a reusable component to present Smart Suggestions in the UI. Smart Suggestions themselves are generated on demand and supported by a microservice that is written in Python. This microservice takes the user input and the task name as input. We used OpenAI's ChatGPT using the \textit{GPT3.5-Turbo model}~\cite{openai2023gpt4}. We use a zero-shot prompting approach, meaning that no example data is provided in the prompt. Furthermore, the required context is included within the prompt (e.g., article title and abstract) to ensure sufficient information is available to fulfill the task. A system message explains to the LLM what task to perform. Afterwards, a user message inputs the relevant contextual data. Finally, a function calling~\cite{chatgpt-function-calling} provides the LLM with the specification of the desired output. The function calling ensures that data is returned in a machine-readable format (JSON) and follows a specification that can be parsed by the frontend. For some prompts, we explicitly instructed the LLM to provide the JSON format in the system prompt, which slightly improved the reliability of the expected output format. 

\section{Evaluation}
We conducted a user study to evaluate both the approach and implementation. Regarding the approach, our aim is to evaluate how relevant the individual guidelines are. For the implementation, we want to assess the usefulness of the Smart Suggestions when implemented in the UI. We will now discuss the evaluation in more detail. 

\subsection{Demographics and Setup} 
To evaluate our approach, we recruited participants that are experienced users of the ORKG, but are not part of the development team, and therefore represent real-world users. At the time of the evaluation, the participants were part of a grant program where they received a monthly monetary reward for curating ORKG content for a period of six months. We deemed these participants to be most suitable to participate in our study, as they are able to determine what would help them to speed up the process of content creation and curation. In total, we recruited six ORKG curators to be part of the study. Five of them have a PhD degree as their highest level of education and one of them has a master degree. Three of them have Computer Science as their field of expertise, and the remaining ones are affiliated with Earth Science, Transport and Traffic Science, and Hydraulic Engineering. 

The evaluation is conducted by means of an online questionnaire that is divided into three parts. The first part evaluates the approach, as presented in \autoref{tab:approach}, by determining the importance of the individual guidelines according to the participants. For each of the guidelines, they are asked to rate them on a Likert scale, ranging from 1 ``not at all important'' to 5 ``very important''. The second part presents the six use cases as discussed in \autoref{section:use-cases}. For each of the use cases, four different domain examples are displayed via screenshots of the actual UI and response from the LLM. The use cases comprise a generic example (i.e., without any data), COVID-19 from epidemiology, global sea level rise from climate science, and women's mental health in academics from gender studies. For each example, the participants are asked to rate the correctness and usefulness, again using a Likert scale, ranging from ``strongly disagree'' to ``strongly agree''. In the final part of the evaluation, participants are asked to answer eleven questions regarding their attitudes towards Smart Suggestions and ChatGPT in general. Additionally, an open question gave participants the ability to provide feedback about additional Smart Suggestions use cases in the ORKG UI. And the last open question gave participants the possibility to provide other remarks.

\subsection{Results}
We present and discuss the results of the evaluation in three parts. Firstly, the approach evaluation. Secondly, the use case evaluation. Finally, the evaluation of the participants' attitudes towards Smart Suggestions and ChatGPT. 

\subsubsection{Approach Evaluation}
\begin{figure*}[t]
    \centering 
    \includegraphics[width=0.9\linewidth]{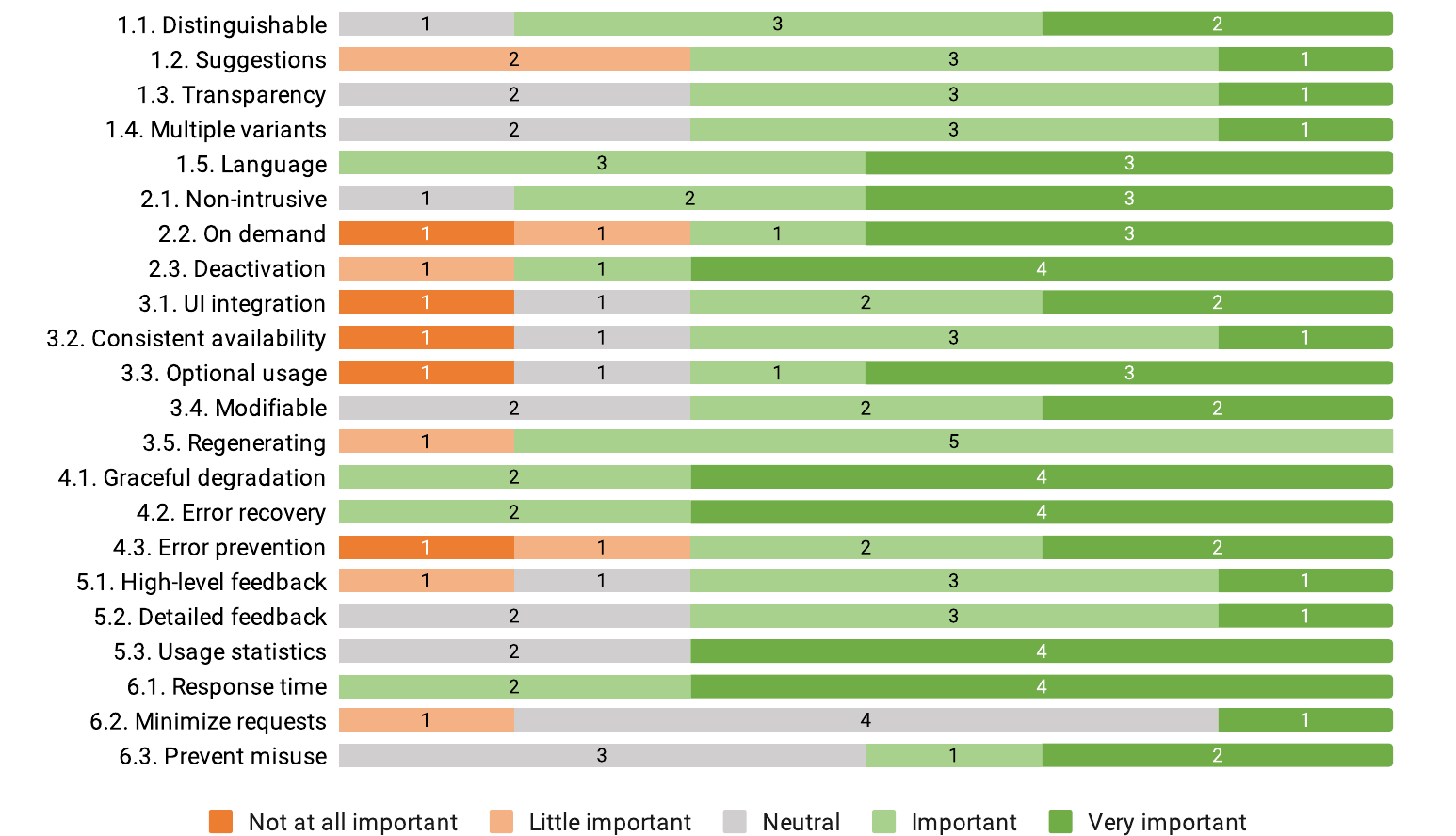} 
    \caption{Approach evaluation results. The individual guidelines are evaluated by means of importance according to the participants. }
    \Description{This figure shows a bar chart for each of the 22 guidelines. The guidelines are scored with a Likert scale, ranging from 'Not at all important' to 'Very important'}
    \label{fig:evaluation-approach}
\end{figure*}

The results regarding the participants' opinions are displayed in \autoref{fig:evaluation-approach}. As can be observed, on average the guidelines are considered to be important, and thus relevant, by the majority of the participants. Participants find guidelines 4.1, 4.2, and 6.1 to be most important. The former two guidelines are related to error management and ensuring that the system remains working. It is therefore expected that participants consider these guidelines to be important. Guideline 6.1 relates to the response time which should be short. Especially when the number of tokens increases (both for input and output), the response time can grow significantly. Therefore, it is crucial to keep the token sizes to a minimum to ensure the response time is within seconds. Participants find guideline 6.2 to be the least important. This can be partially explained by them being potential users of Smart Suggestions and therefore less concerned with the monetary costs that are associated with using LLMs. 

\begin{figure*}[t]
    \includegraphics[width=0.55\linewidth]{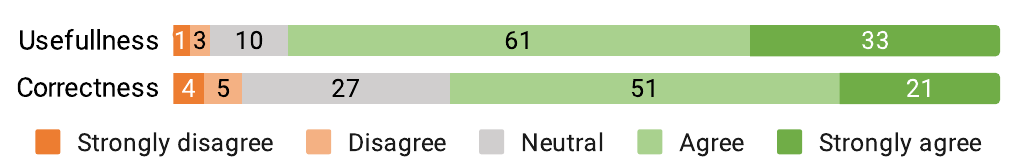} 
     \Description{This figure shows two bar charts for all six use cases. The bar charts show the perceived usefulness and the correctness of the use cases}
    \caption{Results of the usefulness and correctness aggregated for six different use cases and three different domains.}\label{fig:evaluation-use-cases}
\end{figure*}

\subsubsection{Use Case Evaluation}
The use case evaluation results are displayed in \autoref{fig:evaluation-use-cases}. The results are aggregated from all six use cases and the individual domain examples. In general, the perceived usefulness is high which exceeded our expectations. An interesting observation is that the perceived usefulness is higher than the correctness. This can be explained in light of our assumption that Smart Suggestions can provide inspiration to the participants even if the suggestions are not fully correct. This is also in accordance with the evaluation results from the attitudes evaluation.  

\subsubsection{Attitudes Evaluation}
\begin{figure*}[t]
    \centering 
    \includegraphics[width=\linewidth]{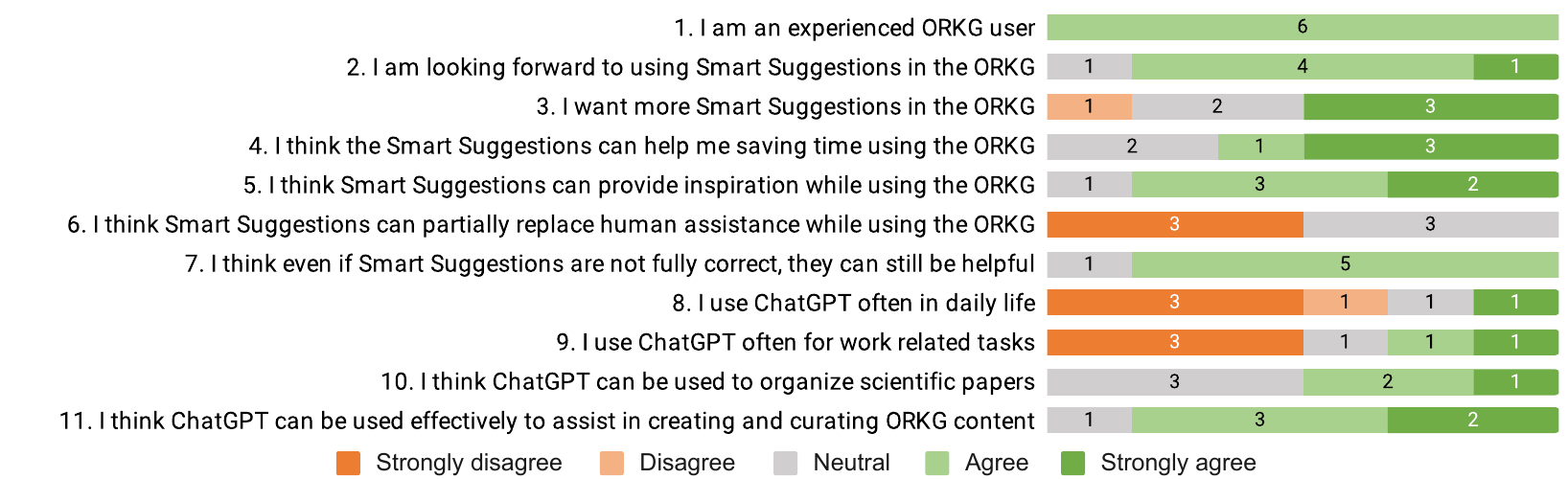} 
    \caption{Evaluation results of the participants' attitudes towards the Smart Suggestions as implemented in the ORKG. Additionally, it shows their opinions and experiences using ChatGPT.}
     \Description{This figure shows bar charts for eleven additional questions to determine the participants attitudes towards our approach. Each answer is provided by a Likert scale ranging from 'Strongly disagree' to 'Strongly agree'}
    \label{fig:evaluation-attitudes}
\end{figure*}

Finally, the results of the participants' attitudes evaluation are displayed in \autoref{fig:evaluation-attitudes}. As question 1 indicates, the participants indeed consider themselves to be experienced with the ORKG, which was our target when recruiting participants. Overall, the results indicate that the participants' attitudes are relatively positive and that they appreciate the Smart Suggestions, as indicated by questions 2 to 5. However, question 6 shows that participants think that Smart Suggestions are not able to replace human assistance while using the system. This could either mean that more Smart Suggestions are needed (also partially confirmed by question 3), or Smart Suggestions themselves are not sufficiently able to provide the same assistance as humans can provide. Questions 8 and 9 indicate that most of the participants are not using ChatGPT regularly in their daily lives, which means that participants were most likely less opinionated towards LLM usage in general. However, they do believe that ChatGPT can be successfully leveraged to organize science (question 10) and can be used effectively in the ORKG (question 11). 

\section{Discussion} 
With the rise of LLMs, their presence in everyday tasks and applications is growing. However, in order to integrate LLMs into existing workflows, one has to be sure the LLM provides correct and helpful information. We applied our implementation to the scholarly knowledge organization domain, where data correctness and trustworthiness are key parts of the system. If the scholarly knowledge graph contains incorrect knowledge, it loses its potential to become a valuable asset for researchers. Therefore, integrating LLMs in this domain, with all their potential weaknesses, such as hallucinations or difficult-to-trace references, is challenging. We specifically address those weaknesses by deliberately making LLMs a non-vital part of this workflow. We decided to integrate the LLM in an assistive manner, by providing guidance to users. This largely mitigates the previously mentioned weaknesses of LLMs, as humans are verifying the correctness of the suggestions before any data is added to the knowledge graph. 

\textit{Prompt Engineering Experiences and Challenges.}
Some tasks performed by LLMs are more suitable than others. A seemingly simple task of verifying whether a piece of text follows certain rules (e.g., not capitalized, short, reusable, etc.) turns out to be rather difficult for LLMs. Ideally, the response of an LLM would be a list of true/false values indicating whether a specific rule is met. In our experience, the responses of providing true/false results were highly variable and therefore too unreliable to be usable in a real-world environment. Sending the identical prompt multiple times resulted in opposite answers, even when the LLM was configured to reduce variations of responses (by setting the model temperature to zero). Also, asking for a percentage instead of a true/value value turned out to be too unreliable. Finally, we decided to remove the evaluation score altogether. The prompts were modified to ask for feedback instead of a scoring evaluation of an input. Even though answers from the LLM may still differ significantly for the same input, when open feedback is expected, this is acceptable, or even desirable. This enables users to get additional feedback every time a request is made. To conclude, tasks that provide specific structured evaluations are less suitable for LLMs than tasks that provide open-ended feedback to users. This is an important aspect of fostering user adoption and managing user expectation. 

Another challenge related to non-conversation LLM usage is retrieving responses in a machine-readable format. Initially, we added sentences such as ``return in JSON'' to get the desired format. In our experience, this method was unreliable. Not only was the response frequently not clean JSON but mixed with text (e.g. the response: ``Sure, I can do that. [JSON]''), also responses were often not parsing due to syntax errors. Instead, we used ChatGPT's ``function callings'' which improved the reliability of the response parsing significantly. This enabled us to remove the logic from the microservice that performs fuzzy parsing of responses. Although this method also occasionally returns non-parsable formats, due to the error recovery options in the UI, this results in only a minor inconvenience for users. Finally, we further optimized our prompts using a trial and error method, ensuring the LLM was returning the desired response. 

\textit{Generalizability.}
In our work, we used the scholarly domain as an example of a challenging application domain, but the approach is applicable outside this specific domain as well. The guidelines in \autoref{tab:approach} are presented in such a way that no assumptions are made about the domain. By presenting the guidelines as requirements, we facilitate adoption of the guidelines when integrating LLMs in existing UIs. The scholarly domain poses several challenging factors that are also applicable in other domains, such as correctness and trustworthiness of data. By presenting LLM responses as suggestions, and having manual verification of data by users, we propose a rather conservative approach for LLM usage in existing UIs. However, this conservative approach is easier to adopt in real-world systems, as LLM performance is less critical when LLMs are employed in a suggestive manner. We believe the guidelines offer a practical resource for the HCI community seeking to adopt the latest technologies without requiring significant changes in existing workflows and keeping the weaknesses of LLMs in mind. The use cases in which LLM supported can be integrated differs depending on the domain. The six use cases from our work are domain-specific. However, LLMs have already proven to be helpful in a plethora of other domains, including education, customer service, legal, etc. The integration of our approach for those domains is possible after domain-specific use cases are identified. Finally, the previously discussed LLM experiences are domain agnostic and provide insights for LLM integration in UIs in general. 

\textit{Limitations and Future Work.}
The evaluation is conducted with a small number of participants, and it is therefore not possible to draw firm conclusions from the evaluation results. The relatively low participant number can be explained based on the selection criteria for our sample. The selected participants all had at least several months of experience using the ORKG and could therefore provide informed feedback about our approach. We deliberately excluded ORKG system developers from our study to prevent bias towards the platform. Additionally, we did not include inexperienced users to prevent bias caused by lack of experience. Despite the small number of participants, we consider the evaluation results to be relevant as it provides early feedback to further guide this research its development. We plan to continue the evaluation of our approach in a real-world setting. In this setting, actual system users interact with the Smart Suggestions and are thus better able assess the correctness and usefulness of the approach. Also, we hope to collect feedback to guide further prompt engineering efforts and to identify additional use cases where Smart Suggestions can be employed. 

In our implementation, we only considered a single LLM as detailed in \autoref{section:implementation-technical-details}. LLMs are varying widely, especially when fine-tuning LLMs (i.e., training them for a specific task). Therefore, some of the previously mentioned experiences might be less applicable to models other than GPT3.5-Turbo. Keeping this in mind, we tried to focus on experiences that are relevant to LLM usage in general. In future work, we plan to evaluate different open source models that are pre-trained to perform specific use case tasks. In addition to the open source benefit (i.e., transparency), less resources (i.e., computational and monetary) are consumed by smaller models.

\section{Conclusion}
In this work, we presented an approach that provides guidelines for non-intrusive integration of Large Language Models (LLMs) into existing User Interfaces (UIs). These guidelines present LLM integration as a support system for users. Some of the key elements of the guidelines include optional and on-demand usage, transparency, and distinguishability. Mainly due to these guidelines, it becomes possible to integrate LLMs into systems, even if the results of the LLMs are not always accurate. We used the guidelines to implement Smart Suggestions in the scholarly knowledge graph UI. The Smart Suggestions are designed to support users in knowledge graph curation and population. Afterwards, we conducted an evaluation, consisting of expert users, that evaluated both the approach and implementation. The evaluation results indicated that the participants were positive about the usefulness of the presented use cases. Furthermore, the participants confirmed that LLM support can be helpful, even if the suggestions are not fully accurate. In future work, we will continue with the development of Smart Suggestions, extending the implemented use cases and performing larger-scale evaluations based on real-world usage. 

\begin{acks} 
This work was co-funded by the European Research Council for the project ScienceGRAPH (ID:~819536), by NFDI4DataScience (ID:~460234259), and by the TIB Leibniz Information Centre for Science and Technology. We want to thank the entire ORKG team for their contributions to the ORKG platform, including research and development efforts.
\end{acks}

\bibliographystyle{ACM-Reference-Format}
\bibliography{refs}

\clearpage
\appendix
\section{Use Cases and Prompts}
\label{appendix:use-cases}
\setcounter{table}{0}
\renewcommand{\thetable}{A\arabic{table}}
\begin{table*}[b]
\centering
\caption{Six implemented use cases within the ORKG, grouped by two types: Closed Recommendations and Open Feedback. For each use case, the LLM prompt is listed, consisting of a system and user prompt. }
\label{tab:use-cases}
\resizebox{.95\textwidth}{!}{%
\begin{tabular}{@{}p{2.9cm}|p{7cm}|p{7cm}@{}}
\toprule
\textbf{Use case} & \textbf{Description} & \textbf{Prompt}\\ \midrule
1. Related Predicates & When making statements in an RDF knowledge graph, a subject, predicate, and object are required. The object can either be a resource (a piece of information with an identifier that can be linked to) or a literal (information that cannot be linked to, such as a string, numbers, dates, etc.). This type of Smart Suggestion recommends predicates to users based on a set of predicates coming from the existing paper description. & \textbf{System prompt:} You are an assistant for building a knowledge graph for science. Your task is to recommend additional related predicates based on the set of existing predicates. Recommend a list maximum 5 additional predicates.

\textbf{User prompt:} The existing predicates are: [list of predicates]
\\\midrule
2. Related Objects & This relates to the previous task but aims to find a set of related objects instead. Since it requires a prompt that provides the LLM with the necessary context, this is only activated for a selected set of predicates, namely: research problem, method, and approach. Thus, each of these predicates has its own prompt. & \textbf{System prompt:} A [research problem] contains a maximum of approximately 4 words to explain the research task or topic of a paper. Provide a list of maximum 5 research problems based on the title and optionally abstract provided by the user.

\textbf{User prompt:} [paper title] [abstract]
\\\midrule
\multicolumn{3}{@{}p{\textwidth}@{}}{\textbf{Open Feedback}} \\\midrule
3. Literal Applicability & In addition to creating resources at the object position of a statement, RDF also allows creating literals, which resemble a piece of textual information that cannot be linked to. Based on our previous experiences with ORKG users, we learned that it can be difficult for users to decide whether an object should be a resource or a literal. This Smart Suggestion helps to determine the most appropriate type when creating an object. It is evaluating if a piece of text should indeed be a literal, or if it is more appropriate as a resource. & \textbf{System prompt:}
You are an assistant in building a knowledge graph for science. You task is to advice users whether they should use a RDF resource or RDF literal. Based on a user-provided label, advice whether the type should be 'literal' or 'resource'. Literals are generally larger pieces of text and are not reusable, resource are atomic and can be reused. 

\textbf{User prompt:} [label]
\\\midrule
4. Decomposable \newline Resources & If resources are represented in an atomic fashion, they contain general information and can thus be reused more easily. This facilitates interconnections which enhances the graph quality. This use case evaluates if a resource label can be decomposed into multiple labels, or if the content is already sufficiently atomic. & \textbf{System prompt:} You are an assistant for building a knowledge graph for science. Provide advice on if and how to decompose a provided resource label into separate resources. Only provide feedback is decomposing makes sense.

\textbf{User prompt:} [label]
\\\midrule
5. Predicate Reusability & Similar to the resource reusability, generic predicate labels foster predicate reusability. This check evaluates if a predicate label is sufficiently atomic. & \textbf{System prompt:} You are an assistant in building a knowledge graph for science. Provide feedback whether the provided predicate label is generic enough to make it reusable in the graph and explain how to make it more generic. Examples of properties that are not reusable: population in Berlin (because it contains a location), temperature in degrees Celsius (because it contains a unit). 

\textbf{User prompt:} [label]
\\\midrule
6. Comparison \newline Descriptiveness & To increase the findability of data in the graph, various graph artifacts can be described with additional metadata, such as a title and description. One of these artifacts is literature comparisons. 
This use case evaluates whether the provided description for a comparison contains sufficient details, such as the goal and objectives. & \textbf{System prompt:} Provide feedback to a user on how to improve a provided description text. The description text should give information about the objectives and topics of a scientific tabular related work overview. 

\textbf{User prompt:} [description]
\\ \bottomrule
\end{tabular}%
}
\end{table*}

\end{document}